# Magneto-active composites with locally tailored stiffness produced by laser powder bed fusion

Kilian Schäfer*[a,f], Matthias Lutzi[a,f], Muhammad Bilal Khan[a,f], Lukas Schäfer[a,f], Imants Dirba[a,f], Sebastian Bruns[b,f], Iman Valizadeh[c,f], Oliver Weeger[c,f], Claas Hartmann[d], Mario Kupnik[d], Esmaeil Adabifiroozjaei [e], Leopoldo Molina-Luna [e], Konstantin Skokov[a,f], Oliver Gutfleisch[a,f]

[a]Functional Materials, Institute of Materials Science, Technical University Darmstadt, 64287 Darmstadt, Germany

[b]Physical Metallurgy, Institute of Materials Science, Technical University Darmstadt, 64287 Darmstadt, Germany

[c]Cyber-Physical Simulation Group, Technical University of Darmstadt, Dolivostr. 15, Darmstadt 64293, Hessen, Germany

[d]Measurement and Sensor Technology Group, Technical University Darmstadt, 64289 Darmstadt, Germany

[e]Advanced Electron Microscopy, Institute of Material Science, Technical University of Darmstadt, 64287 Darmstadt, Germany

[f]Additive Manufacturing Center, Technical University Darmstadt, 64289 Darmstadt, Germany

Abstract

Additive manufacturing technologies enable the production of complex and bioinspired shapes using magneto-responsive materials, which find diverse applications in soft robotics. Particularly, the development of composites with controlled gradients in mechanical properties offers new prospects for advancements in magneto-active materials. However, achieving such composites with gradients typically involves complex multi-material printing procedures. In this study, a single-step laser powder bed fusion (LPBF) process is proposed that enables precise local adjustments of the mechanical stiffness within magneto-active composites. By utilizing distinct laser parameters in specific regions of a composite containing thermoplastic polyurethane and atomized magnetic powder derived from hard magnetic Nd-Fe-B, the stiffness of the composite can be modified within the range of 2 to 22 MPa. Various magneto-responsive actuators with locally tailored stiffness are fabricated and their magnetic performance is investigated. The enhanced response exhibited by actuators with locally adjusted mechanical properties in comparison to their homogeneous counterparts with identical geometries is shown. As a demonstration of biomedical application, a magnetically responsive stent with localized adjustment is presented with the ability to meet specific requirements in terms of geometry and local stiffness based on an individual's anatomy and disease condition. The proposed method presents an approach for creating functionally graded materials using LPBF, not only for magneto-active materials but also for several other structural and functional materials.

## 1. Introduction

Mechanically soft actuators and sensors enable a safe and compliant interaction between humans and their environment [1,2]. Among many existing efforts to develop such devices, magnetic actuation mechanisms are particularly well-known for the possibility to operate rapidly without wires and in confined spaces [3,4]. Different applications such as soft grippers [5], tactile sensors [6,7], artificial muscles [8,9], compliant wearables [10], prosthetics [11] and biomedical devices [12–14] are currently explored with magneto-active soft materials. Complex and bioinspired shapes of magneto-active composites can be fabricated using additive manufacturing techniques [5,15–19]. However, in order to translate the functional characteristics of biological systems into technology, it becomes imperative to fabricate composites that exhibit localized variations in both mechanical and functional properties [20,21]. It is worth noting that natural materials frequently possess gradients in their mechanical properties [22–28]. In addition, composites with stiffness gradients can advance the development of soft robotic systems, e.g., an exosuit, by reducing the stiffness mismatch at the interface of soft functional materials and rigid components, which can otherwise lead to mechanical failure [29]. It was already shown, that for the mechanical performance of porous composites, a smooth transition between two materials with different stiffnesses outperforms a blunt transtation [30].

In additive manufacturing, achieving localized variations in the mechanical properties of different structural materials is typically accomplished through multi-material printing, aiming to realise a diverse range of stiffness within a given structure. This includes, for example, combining nozzles for different resins [31–34] or filaments [35]. Despite improvements in multi-material additive manufacturing technologies in recent years, challenges remain to be addressed and overcome, including cross-contamination, limited material selection, low production throughput and complex hardware adjustments [36,37]. Recently, the local mechanical property adjustment with a single-step process was demonstrated with vat polymerization technologies [38,39]. Gradients in mechanical properties were produced by controlling the degree of polymerization with different light intensities. However, the production capabilities of magneto-active composites with vat polymerization technologies are limited since the magnetic particles in the resins affect the polymerization processes by absorbing the incident radiation [40]. Thus, the inclusion of magnetic particles is typically restricted to a range of 6 - 8 weight % (wt. %) due to their adverse impact on resin viscosity and substantial reduction in optical transmissivity [41]. Since the deformation and forces of magneto-responsive materials scale with the magnetic moment per volume, a high magnetic filler fraction is usually desired [3]. For a uniformly magnetized beam, the deflection is maximized at around 20 volume % (vol.%), which is around 70 wt.% and the energy density is maximixed at around 30 vol. %, which is around 80 wt.% [3]. It was already shown that with laser powder bed fusion (LPBF) of polymer-metal composites a high magnetic filler fraction of up to 93 wt.% is possible [42,43].

In this study, an approach is presented for modifying the stiffness of a magneto-active composite at specific local regions. The proposed method utilizes a single-step laser powder bed fusion (LPBF) process. The composite consists of hard magnetic Nd-Fe-B particles within a thermoplastic polyurethane (TPU) matrix. The LPBF process involves selectively fusing layers of powder using a laser beam [44]. By adjusting the laser parameters at distinct areas during the process, it becomes possible to locally modify the mechanical properties of the composite. After a verification of the filler fraction, the density and magnetic performance of the composites as a function of the laser parameter is investigated. The range of achievable mechanical property modifications is then investigated through tensile tests conducted on composites produced with varying laser parameters. To validate the presence of stiffness gradients within a single sample, line scans of Vickers indentations are performed using nanoindentation. Different actuator shapes are presented where the stiffness is locally adjusted, resulting in an enhanced response to magnetic actuation fields. This advancement opens up possibilities for manufacturing magneto-active materials customized for individualized applications and facilitates the integration of soft functional components with rigid components in robotic systems.

## 2. Materials and Methods

### 2.1 Materials

Magneto-active composites were produced using the thermoplastic polyurethane (TPU) based polymer "Flexa Grey" from *Sinterit* (*Poland*) in combination with the spherical hard magnetic powder "MQP-S" which is based on the $Nd_2Fe_{14}B$ system from *Neo Magnequench* (*Singapore*). TPU-based polymers are composed of block copolymers containing alternating rigid and flexible segments. The rigid crystalline segments serve as crosslinks, while the flexible amorphous segments contribute to the flexibility. [45]. TPU-based polymers possess excellent mechanical properties, such as high ductility and abrasion resistance, making them well-suited for soft robotics applications. Moreover, they can be made biocompatible for medical devices [3,46]. These polymers are melt-processable due to the behavior of the rigid segments becoming disentangled at elevated temperatures [3]. Therefore, they can be processed with LPBF as well, as already has been shown in the literature [46–50].

The properties of the initial powders, as provided by the literature [51,52], are given in Table 1. Since the manufacturer only provides the printout density of the Flexa Grey material, the theoretical density of the Flexa Grey polymer was determined by fully melting the powder in cylinder with a radius of 4 mm and height of 16 mm. Then the mass of the fully molten polymer was determined. The remanence $J_r$ is the remaining polarization of the material after removal of an external field. The coercive field strength describes the resistance of the material against demagnetization. The size distribution and morphology of the powders are presented in Figure 1 a) for Flexa Grey and b) for the MQP-S. The spherical morphology of the MQP-S powder results in excellent flowability and therefore good processability by LPBF. The hysteresis loop and the initial magnetization curve of the hard magnetic MQP-S powder are presented in Figure 1 d). The magnetic powder is an exchange-coupled material, where the remanence is isotropically increased due to an effective coupling (compare Figure 1 d)) of a magnetically soft and hard nano-zisized grains within the larger particle (compare Figure 1 c)) [53]. A bright-field scanning transmission electron microscopy (STEM) is presented in the insert, where the nanocrystalline main phase and alpha iron could be detected using energy dispersive spectroscopy (EDS). A high remanence is beneficial for magnetically actuated soft robotic applications, since the forces scale with the magnetic moment of the composites [3]. A high loading fraction is not always possible and results often in limitations of the processability and in an increase of the part porosity [54]. Due to the spherical morphology and the high isotropic remanence, the MQP-S powder is well suited for the LPBF of magnetic soft robotics components. To produce the composites, the magnetically isotropic powder was used in the thermally demagnetized state, as received from the manufacturer.

For structural and chemical investigations a thin lamellae of from the MQP-S powder was prepared using focused ion beam lift-out technique (FIB: JEOL JIB-4600F), and the lamellae were studied using transmission electron microscope (TEM: JEM-2100, JEOL) in both transmission and scanning transmission modes. The machine was equipped with an Oxford X-Max80 energy dispersive X-ray spectroscopy (EDX) detector, thus chemical analysis was also done in scanning transmission mode.

*Table 1: Feedstock properties of the powders as given by the manufacturer. *The theoretical density of the Flexa Grey polymer experimentally obtained since it not provided by the manufacturer. ** Range is given by manufacturer, the measured value of the used powder is 0.88 T.*

| Powder | Material | Particle size (μm) | Theoretical density (g/cm³) | Remanence $J_r$ (T) | Coercivity $\mu_0 H_c$ (T) |
|---|---|---|---|---|---|
| Flexa Grey *Sinterit* [51] | Matrix | 20 - 105 | 1.09* | n/a | n/a |
| MQP-S *Neo-Magnequench* [52] | Filler | 35-55 | 7.43 | 0.75 | 0.84 – 0.94** |

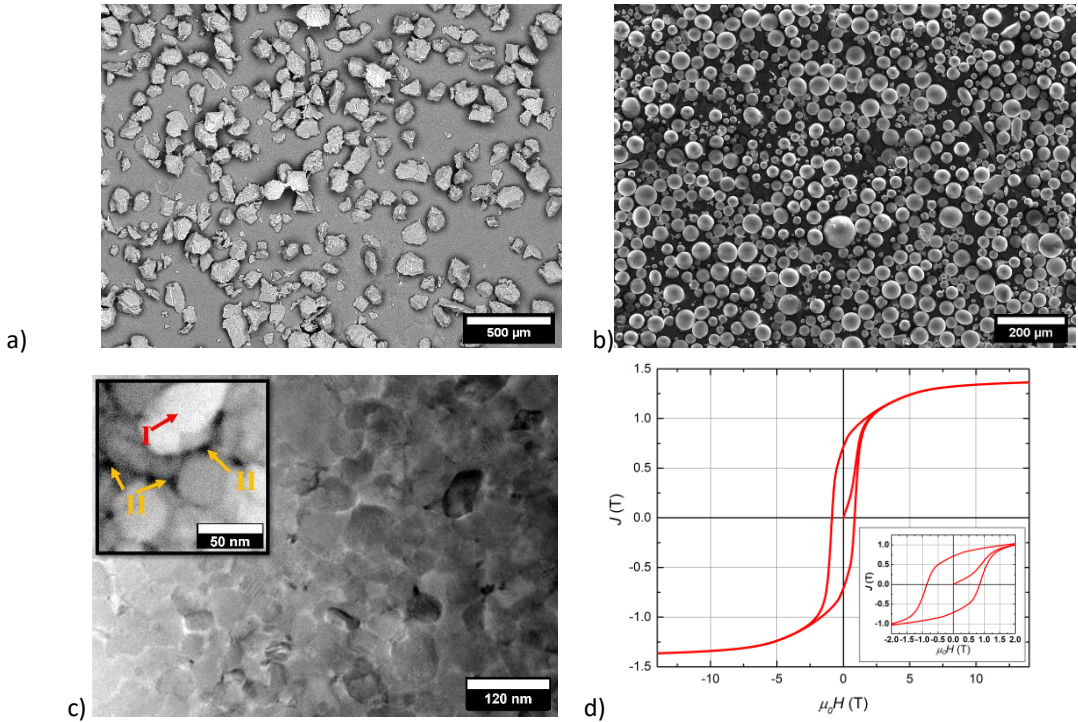

*Figure 1: SE-SEM images of the initial powders a) TPU based Flexa Grey from Sinterit and b) spherical MQP-S based on NdFeB from Neo Magnequnech. c) Transmission electron microscope (TEM) cross-section image of the nano-sized grains within a single MQP-S particle. Insert: Bright-field STEM image of the nanostructure in one MQP-S particle. Local composition was measured with EDX. Area I has the composition $Fe_{88.4}Nd_{11.6}$, while the area II is more iron, rich: $Fe_{95.3}Nd_{4.7}$. The analysis indicates $Nd_2Fe_{14}B$ as main phase and an iron righ phase, probably alpha iron as secondary phase. d) Hysteresis loop of the magnetic powder MQP-S with the initial magnetization curve measured at room temperature.*

The feedstock for the LPBF was produced by mixing the two powders with the shaking mixer from *TURBULA®* (*WAB-GROUP*, Switzerland) for 30 min. The nominal filler fraction $\omega_f$ was 50 weight % (wt. %) which corresponds to a volume fraction $\phi_f$ of around 12.7 %. The conversion was calculated with the following equation:

$$\phi_f = \frac{\omega_f/\rho_f}{\frac{(1-\omega_f)}{\rho_m} + \omega_f/\rho_f}$$

Where $\rho_f$ is the density of the filler (MQP-S) and $\rho_m$ the density of the polymer matrix.

The filler fractions of the produced samples were verified by thermogravimetric analysis (TGA). An *STA 409 CD Simultaneous thermal analyzer* (*Netzsch,* Germany) was used with a heating rate of 10 °C/min. The experiment was performed under nitrogen atmosphere and a temperature ranging from 30 °C to 600 °C.

## 2.2 Manufacturing

For the LPBF process, the commercial device "Lisa Pro" from *Sinterit (Poland)* was used which is equipped with a continuous-wave mode 5 W (λ = 808 nm) infra-red (IR) laser diode. The powder bed is heated by resistive heaters whereas the powder surface is heated with infrared lights. The print chamber has the dimensions 150 x 200 x 260 mm. The laser scanning direction is rotated by 90° for each layer. The process parameters are listed in Table 2:

*Table 2: Process parameters of the Lisa Pro for the Flexa Grey material.*

| Parameter | Powder surface temperature (°C) | Print chamber temperature (°C) | Laser scanning speed (m/s) | Laser spot size (mm) | Hatch distance (mm) | Layer height (mm) |
|---|---|---|---|---|---|---|
| Value | 102 | 70 | 0.054 | 0.4 | 0.316 | 0.125 |

For all samples these parameters were held constant, while only an additional parameter, the so-called energy scale, which is a dimensionless parameter that defines how much energy is transferred to the powder bed with the laser, was adjusted to produce samples with different mechanical properties. The parameter was varied from 0.6 to 3.0 in steps of 0.4 (0.6; 1.0; 1.4; 1.8; 2.2; 2.6; 3.0). For a lower energy scale lower then 0.6, the samples showed a low mechanical integrity and for values above 3.0, the samples exhibit geometrical distortions.

The Sinterit software allows the assignment of one parameter set per digital file, for example, an STL (Standard Tessellation Language) file. To realize locally tailored properties, the energy scale parameter is varied locally. For the process, the digital model of the sample needs to be separated into distinct regions representing the desired varied properties. In situations where a gradient in the x-y direction is required, areas featuring different process parameters can be positioned adjacently to one another [Figure 2 (a)]. However, the software automatically produces a laser scanning strategy with an outline and an infill, as captured during the process shown in Figure 2 c). Since the outlines are produced with lower laser power, these regions have a lower mechanical cohesion. This can be avoided by selecting the thickness of these regions to be the process parameter hatch distance (h), which is the separation between two consecutive laser scans. If the regions have the same dimension as the hatch distance in the direction, where a mechanical property gradient is needed, a gradual change of the process parameters can be achieved without the formation of outlines [Figure 2 (b)]. The powder surface during the LPBF process after scanning with this strategy is shown in Figure 2 d). If the gradual change in mechanical properties is needed in the z-direction, i.e., the build-up direction, the dimensions, represented by 'l', of the regions in the z-direction should be a whole number multiple of the process parameter layer height [Figure 2 (e)]. The different regions can have different dimensions in the z-direction if they satisfy this condition. Otherwise, the Sinterit software would produce empty layers in between the different regions.

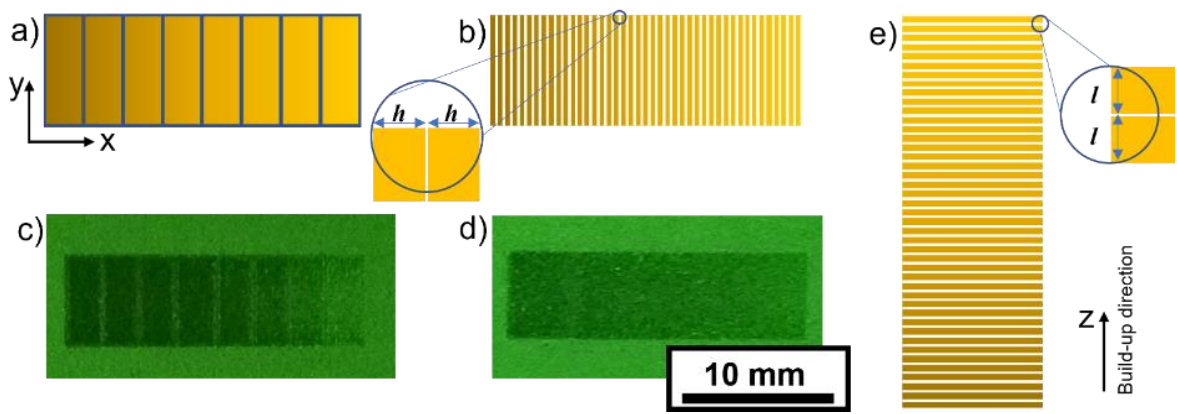

*Figure 2: Schematic image to illustrate how to divide the digital models to produce parts with locally assigned process parameters. In the x-y plane, the size of the regions should have the same thickness as the process parameter hatch distance h, as shown in b) and captured during the process in d). If the thickness is larger than the hatch distance, the Sinterit software will automatically create outlines and infills, as shown in a) and captured during the process in c). The division of the digital file in z-direction is shown in e), where the height l of the digital parts must be an integer multiplier of the process parameter layer height.*

**2.3 Characterization**

The mechanical properties of the composites were investigated with tensile and nanoindentation tests and the magnetic properties were studied with a pulsed field magnetometer. The actuator shapes were magnetized in a pulsed field magnetometer and their actuation behavior was investigated with an electromagnet. Details about the sample geometries and analysis methods are provided in this section.

Different geometries were produced by LPBF. Three cubes with a side length of 7.5 mm were produced for each energy scale parameter for the determination of the geometrical density and the magnetic properties. To visualize the microstructure of the composites, light microscopy was done on the cubes which were produced with 0.6 and 3.0 energy scale process parameter. For the mechanical characterization, five dumbbell-shaped samples for each energy scale parameter according to the ASTM D638 standard with sample type IV were produced. In order to reduce the powder consumption, the size of the mechanical tensile test geometries was reduced to 50 % compared to the standard. The orientation of the parts in relation to the build-up direction has an influence of the properties of TPU produced with LPBF [47]. The long axis of the samples was parallel to the built-up direction of the LPBF device. This direction was chosen, to match the compression direction of the artificial muscles.

Samples with a linear increase in energy scale from 0.6 to 3.0 were produced to verify a mechanical gradient within a sample. Two square cuboid samples with a side length of 10 mm were produced vertically with a height of 20.25 mm, consisting of 54 segments with a layer height of 375 µm (3x "layer height"). The dimensionless energy scale parameter was varied linearly in the digital segments from 0.6 to 3.0 in steps of 0.05. In one sample the energy scale was linearly increasing with the build-up direction and in the other one decreasing with the build-up direction. Another sample was produced with a linear gradient in the energy scale, but with the gradient in the horizontal direction, perpendicular to the build-up direction. Here, the digital segments had a thickness of 316 µm (hatch distance). In total 72 segments were used, which resulted in the total dimensions of 10x10x22.75 mm.

Also, magneto-responsive actuators were produced with different shapes. Foldable cubes, consisting of four 10x10x1 mm plates connected with 2.5 mm segments with a thickness of 0.5 mm. Here, cubes with a homogenous energy scale parameter of 3.0 were produced as well as cubes where the platforms were produced with 3.0 and the smaller segments with 1.0. The foldable cubes were magnetized in their folded state within a pulsed field magnetometer (*HyMpulse* from *Metis Instruments*, Belgium) with 6.85 T. Two different artificial muscle shapes were produced with an upright printing position and a length of 20 mm. The detailed shape and dimensions are depicted in Figure A 1 and Figure A 2. These shapes were magnetized in a compressed state with 16 mm length with the pulsed field magnetometer. A stent shape was produced as well with homogenous parameters and with locally adjusted parameters. The dimensions of the stent are shown in Figure A 3.

The tensile tests were performed at room temperature and according to the ASTM D638 standard for five samples per laser parameter. An *Inspekt table 5* from *Hegewald and Peschke* (Germany) with a 100 N load cell and a test speed of 5 mm/min was used. The displacement of the traverse was used to evaluate the sample strain. The distance between the grips was 32.5 mm (50 % of ASTM D638 dimension).

The magnetic properties of the composites produced with different laser parameters were evaluated at room temperature with external fields of 6.85 T with the HyMpulse pulsed field magnetometer. For the demagnetization correction, a value of $N = ⅓$ was used due to the spherical morphology of the magnetic powder. An isothermal magnetization measurement at room temperature, with an external magnetic field up to 14 T, of the pure MQP-S powder was carried out using PPMS-VSM (Quantum Design PPMS-14). For the measurement the MQP-S particles were embedded in an epoxy to immobilize the particles.

The hardness gradient within each sample was measured via nanoindentation testing using a G200 nanoindenter from *KLA instruments* (USA) equipped with a Vickers geometry indenter tip. The instrument was operated in constant strain rate mode with a 0.05 $s^{-1}$ target strain rate. Vickers indentations of 20 µm depth were placed every 500 µm over 20 mm. Since adequate surface finish is required for these tests, the specimens were ground and polished with up to 4000 grit.

The actuator geometries were magnetized with a 6.85 T pulse with the HyMpulse pulsed field magnetometer. The actuation tests were then performed between the poles of the water-cooled electromagnet (model 3474-140 from *GMW, USA*). The setup is shown in Figure A 5 with the corresponding magnetic field distribution in the different directions. The applied magnetic field strength was measured with a Hall sensor which is attached to one of the pole shoes. The deformation was analyzed optically with the software *ImageJ*.

The imprinted magnetization profile of the cubes was measured with a custom build setup consisting of linear motors and a 3MTS Teslameter. The sensor and linear motors were controlled with the LabVIEW software (*National Instruments, USA*). The Hall sensor moves above the sample with 1 mm steps and then the components of the magnetic field produced by the magnetization of the sample are measured with a spatial resolution of 150 µm at each measurement point.

**2.4 Simulation**

For the magneto-mechanical modeling of the artificial muscles numerical simulations were carried out using the commercial software Ansys (Ansys® Academic Research Mechanical, Release 2021R2). The force on the top surface of the artificial muscles was measured and used as a boundary condition in the software. In addition, a zero displacement boundary condition was applied on the bottom of the artificial muscles, and the plates' surfaces are constrained in the orthogonal direction to the surfaces. The fabricated polymer composites exhibit a nonlinear stress-strain relation and experience finite deformations [55]. Hence, a Mooney-Rivlin three-parameter hyperelastic constitutive model was fitted to the experimental data from Figure 7 and utilized for the simulations [38]. The artificial muscles shown in 9 a) and b) were meshed with the tetrahedral quadratic elements. A convergence study was performed for both geometries, based on which a mesh with 1,097,962 nodes and 504,350 elements was used for the artificial muscle in Figure 9 a) and a mesh consisting of 303,908 nodes and 174,430 elements was employed for the artificial muscle in Figure 9 b).

## 3. Results and discussion

### 3.1. Verification of filler fraction

The filler fraction of magneto-responsive composites is a crucial parameter influencing the magnetic and mechanical properties of the material and influences the resulting forces in an magnetic actuator [3]. The nominal filler fraction is verified by TGA and pure Flexa polymer was measured as a reference. The results of the TGA experiment are presented in Figure 3 as mass percentages of the pure TPU polymer and two composites, produced with different energy scales (0.6 and 3.0). At approximately 300°C, a reduction in mass becomes apparent, indicating the decomposition of the TPU polymer, consistent with the manufacturer's data sheet. The pure Flexa polymer shows a remaining mass of 9.5%, which can be most probably attributed to carbon black, which has a high melting point of 3550°C. The carbon black is often used to increase the energy absorption of polymer powders by the laser radiation [56,57]. The remaining mass at 550°C for the composites is 58.3% for the sample processed with an energy scale of 0.6 and 57.6% for the sample processed at an energy scale of 3.0. Subtracting the mass of the non decomposed material, a remaining mass filler fraction of the magnetic particles of 48.1% and 48.8% were determined for the composites produced with 0.6 and 3.0 energy scale. This is close to the nominal filler fraction of 50.0% and allows to conclude that the desired filler fraction can be achieved independently from the chosen energy scale between 0.6 and 3.0.

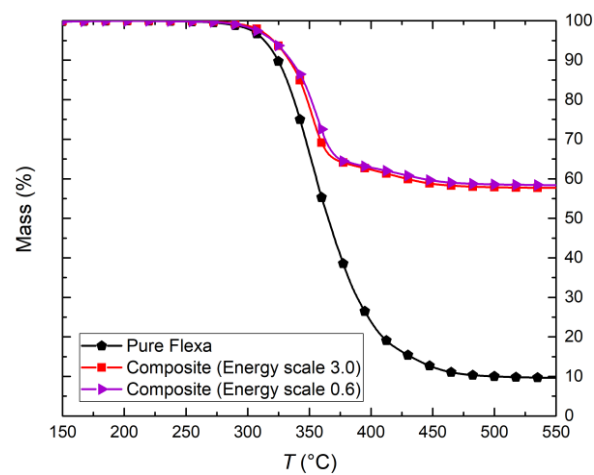

*Figure 3: Mass as function of temperature in TGA measurement of the pure Flexa TPU polymer and the TPU-MQP-S composites produced with different energy scales: 3.0 and 0.6.*

## 3.2 Geometrical density, magnetic and mechanical properties

In order to explore the extent to which material properties can be modified, the density, magnetic characteristics, and mechanical performance of the composite are examined across various energy scale values used during production. The geometric densities of the composites as a function of the energy scale are presented in Figure 4. The average density value of three samples is shown with the corresponding standard deviation. The corresponding values are provided in appendix Table A.1. The theoretical value for a fully dense composite with a filler fraction of 49 wt.% is 1.64 g/cm³, which is indicated by the red line. The porosity is determined with the ratio of the measured density to the theoretical density for a fully dense composite. While the density is increasing from 1.1 to 1.4 g/cm³ with increasing energy scale parameter, the porosity is decreasing from 33 % to 14 %. The different porosity values are visible in Figure A 4 for the composites fabricated with an energy scale of 0.6 and 3.0.

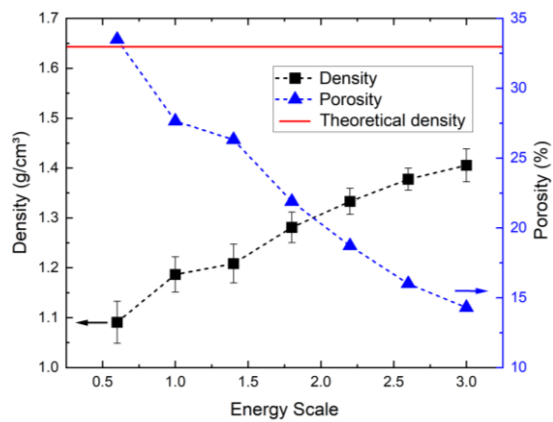

*Figure 4: Geometrical density and calculated porosity as function of the energy scale parameter. The red line at 1.64 g/cm³ indicates the theoretical density for a fully dense compound.*

The magnetic properties of the composites at room temperature are presented in Figure 5, where in the second quadrant demagnetization curves are shown for the composites produced with different energy scale process parameters. They present the polarization J as function of the external field H. The corresponding values are given in the appendix Table A.1. The remanence is increasing with increasing energy scale parameter from 47 mT to 76 mT due to the decreasing porosity. The theoretical remanence value for a fully dense sample with a filler fraction of 48.8 wt. %, which corresponds to 12.27 volume %, is 92 mT. Due to the 14 % porosity of the 3.0 energy scale sample, the remanence should be also 14 % lower in comparison to the theoretical fully dense sample. Therefore, the remanence should be 79 mT, which is very close to the measured value of 76 mT. The coercivity is very similar around 885 mT for the samples produced with higher energy scales and is slightly decreasing for the samples produced with an energy scale of 0.6 and 1.0. The reduced coercivity observed in these samples can be assigned to the weaker mechanical embedding of particles in the polymer matrix caused by the high porosity. This can result in the rotation of inadequately bonded particles during the demagnetization process [54,58]. In this parameter range we consider the nanostructure within the MQP-S particle largely unaffected. It was already reported that the actuation efficiency is decreased when magnetic fillers partially rotate during field application [59]. Therefore, an energy scale parameter of 0.6 can be seen as a lower boundary condition for the process conditions, since samples or areas produced with lower energy scale values will probably result in even weaker mechanically bonded magnetic particles and further decreased torque transmission efficiency. However, the resistance against demagnetization of the composites in the range of the actuation fields of up to 500 mT is sufficient.

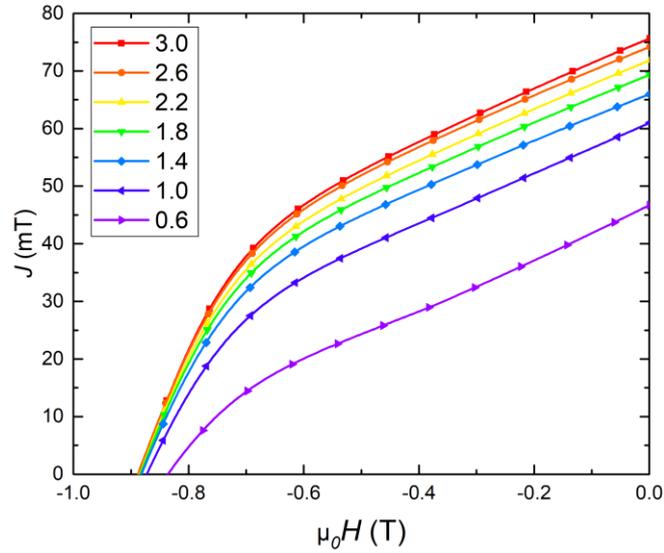

*Figure 5: Second quadrant demagnetization curves where the polarization J is shown as function of the external field H for the TPU-MQP-S composites produced with different energy scale parameters. The samples were magnetized with a pulsed field of 6.85 T.*

To investigate the stiffness range which can be tailored by the process parameters, uniaxial tensile tests are performed. Figure 6 a) shows the engineering stress-strain curves for the samples produced with different energy scale parameters. The corresponding Young's moduli and the ultimate tensile strength $\sigma_m$ are presented in Figure 6 b) with the standard deviation from the five measurements per process parameter. The average mechanical property parameters and the standard deviations of the tensile tests are given in appendix Table A.2.

The Young's modulus and the tensile strength both increase almost linearly with in the range of the energy parameter from 0.6 to 3.0. The stiffness ranges from (2.22 ± 0.40) MPa for an energy scale parameter of 0.6 to (21.99 ± 1.59) MPa for an energy scale parameter of 3.0. The tensile strength is increased from (0.07 ± 0.01) MPa to (0.99 ± 0.07) MPa in the same range. The increase in stiffness and strength is a result of the different amounts of energy deposited in the material during the LPBF process. Insufficient powder melting occurs within the workpieces when using low laser power, resulting in the presence of residual pores in the sintered parts, which can be seen in the porosity values in Figure 4. With increasing laser power, the amount of remaining pores decreases, resulting in a denser workpiece with enhanced stiffness and tensile strength [60]. The tunable stiffness range is smaller in comparison to what can be achieved with multi-material vat polymerization techniques, where, depending on the base material used, e.g., the stiffness ranges from 0.12 to 319 MPa [31].

Referring to the data sheet of the manufacturer, the elastic modulus of the pure TPU is 7.8 MPa, the tensile strength is 3.6 MPa and the strain at break is 210% [51]. Incorporating rigid filler particles results in the reinforcement of the polymer, thereby reducing the deformability of the composite in comparison to the unfilled polymer [3]. The porosity and the weak bonding between polymer and magnetic particles reduces the tensile stength as well. Furthermore, the samples were manufactured in the vertical orientation, implying that the applied tensile stresses are perpendicular to the printed layers. This configuration decreases both the tensile strength and ductility of the samples [47]. The maximum achievable tensile strength at a laser parameter of 3.0 ($\sigma_m(3.0)$ = (0.99 ± 0.07) MPa) is 27.5 % of the manufacturers value and the maximum measured strain at break ($\varepsilon_b(1.8)$ = (44.22 ± 2.87) %) is 20 %. However, this orientation was chosen to investigate the mechanical properties in the same direction as the direction of deformation for the actuation tests with a magnetic field in section 3.4.1.

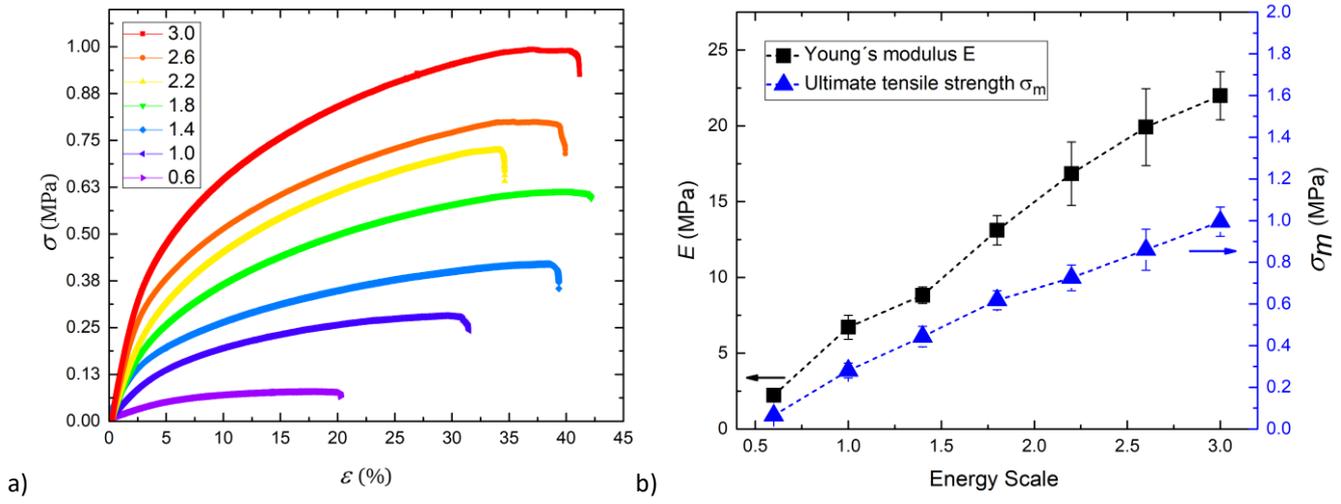

Figure 6: a) Engineering tensile stress strain curves for the samples produced with different energy scale process parameters. b) Young´s modulus as function of the energy scale parameter.

## 3.3 Verification of gradient in mechanical properties

In order to verify the local mechanical property adjustment within one composite sample, nanoindentation can be used [33,61,62]. Samples with different orientation towards the build-up direction were used to test the presented method to locally adjusted mechanical properties. Figure 7 presents the local hardness values of three different composite samples. The values were normalized to the highest value of each sample. The values from the vertical sample with decreasing energy scale from top to bottom are presented Figure 7 a), whereas the values from the vertically produced samples with an increasing energy scale from left to right are presented, and the values from the vertically produced sample with increasing energy scale parameters from top to bottom are shown in Figure 7 b). The horizontally produced sample with increasing energy scale parameters from left to right is depicted in Figure 7 c). The relative change in the average hardness is in the range of 5 to 6 times from the soft to the hard side for all three samples. The edge length of the projected contact area of the Vickers indenter is 99 µm. Due to the mean particle size of 35-55 µm of the metallic particles, it is likely that the local hardness values are strongly influenced if one or more metallic particles are within the contact area. This most likely explains the outliers with the highest hardness values in the data sets.

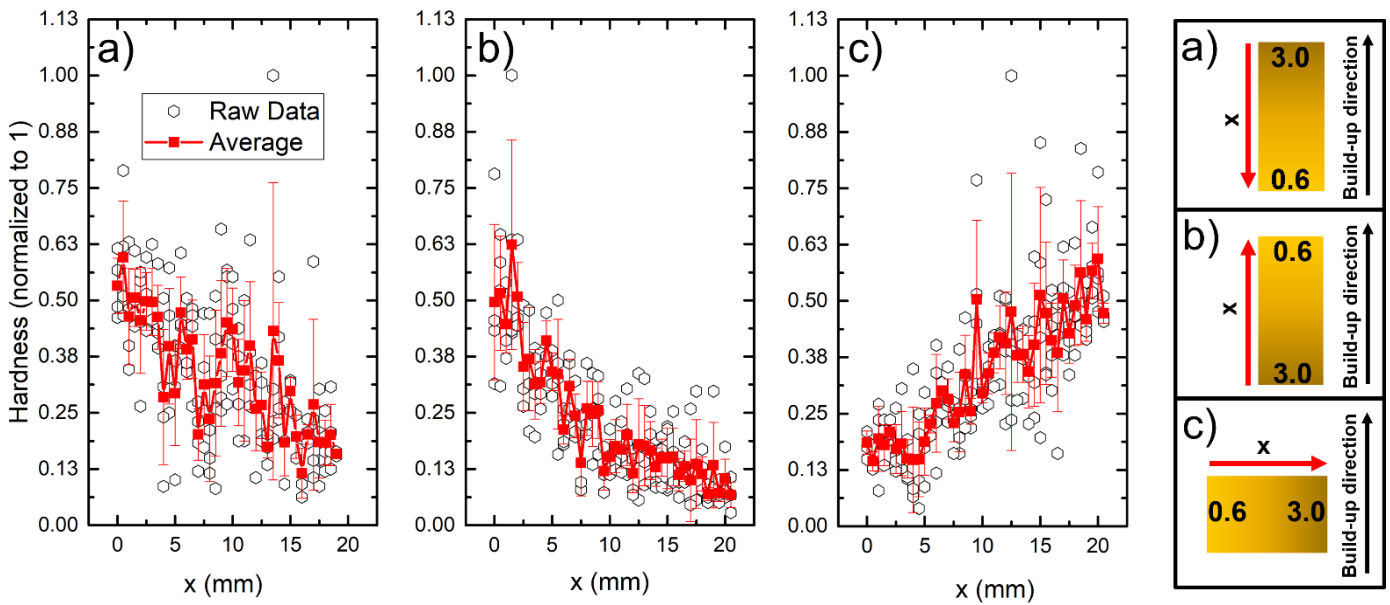

*Figure 7: Normalized hardness as function of the position x on the samples as indicated in the schematic representation on the right side. a) Vertical sample with decreasing energy scale from top to bottom (3.0 - 0.6). b) Vertical sample with increasing energy scale from top to bottom (0.6 - 3.0). c) Horizontal sample with decreasing energy scale from left to right (3.0 - 0.6).*

## 3.4 Magnetic actuation of magneto-active composites with locally adjusted stiffness

In this section, examples are presented where the local property adjustment can be used in soft robotic applications to enhance the actuation performance. Actuators like artificial muscles, foldable cubes and an airway stent are explored to demonstrate that the method is useful for different soft robotics applications.

### 3.4.1 Artificial muscles

First, two designs acting as artificial muscles are investigated with graded and ungraded properties repectively. Two different designs are presented, one which only shows contraction and one which demonstrates contraction and slight rotation. Artificial muscles are actuators that contract, expand or rotate reversibly driven by an external stimulus [63]. Numerial simmulations for uniaxial compression were performed to investigate the deformation behavoir of the two designs and to learn where a local mechanicall property adjustement can increase the actuation performance. The results are presented in Figure 8 a) for the beam shape and b) for the helix shape. The deformation is the largest in the middle of the structures, and, thus, the stiffness should be reduced in this area. The actuators consist of two platforms interconnected by beams, designed to offer flexibility. These platforms can also be connected to other components within a soft robotic system. Using the proposed method, the platforms can be manufactured with a high stiffness that gradually decreases towards the center of the structure, where increased elasticity is required for effective actuation.

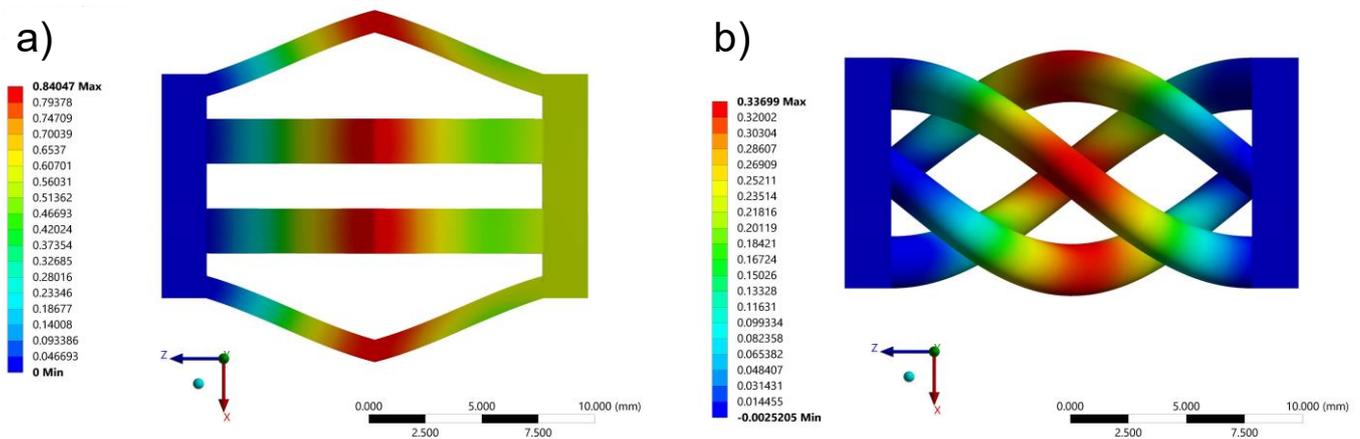

*Figure 8: Result of the numerial simulations, deformation during uniaxial compression. a) Beam shape and b) helix shape.*

As shown in Figure 9, the actuation behaviour of the two actuators with the beam shape are compared: Figure 9 a) with regions which were produced with different energy scale parameters and Figure 9 b) without process parameter variation. The magnetic field direction is indicated by the green arrow. When the magnetic field is increased, the actuators contract since they were magnetized in a contracted state beforehand. The particles experience a magnetic torque that aligns their easy magnetization direction with the external field. This torque is then transferred to the polymer matrix, enabling deformation of the composite under the influence of the external magnetic field [3]. The gradient composite exhibits a deformation of 11.05 ± 1.26% when subjected to a magnetic field of 200 mT, while the non-gradient counterpart only achieves a deformation of 3.24 ± 0.51%. (Figure 11). The deformation of the ungraded design is in agreement with simulation results, which show 3.1% deformation. If the magnetic field is reduced, the actuators reversibly deform back into their original shape. Higher applied actuation fields resulted in a non reversible deformation, therefore the study was limited to application fields of 200 mT.

A similar behaviour is also presented in Figure 10 with the helix shape. Figure 10 a) presents the shape with locally adjusted process parameters to decrease the stiffness locally to allow higher deformation. The sample shown in Figure 10 b) was produced with homogenous process parameters. The deformation as function of the magnetic actuation field ist presented in Figure 11 for the two different shapes with and without locally adjusted properties. The deformation of the helix shape with locally adjusted parameters is larger and a larger rotation can be generated. Here, for a magnetic field of 400 mT, the deformation along the long side of the locally adjusted actuator is 6.52 ± 0.57%, in comparison to the deformation of 2.53 ± 0.54% of the homogenous actuator. Again, the deformation of the homogenenous design is in good agreement with simulation results, which show 2.4% deformation. In addition to contraction deformation, the helix actuator also shows a rotation of the right platform perpendicular towards the long side of the shape.

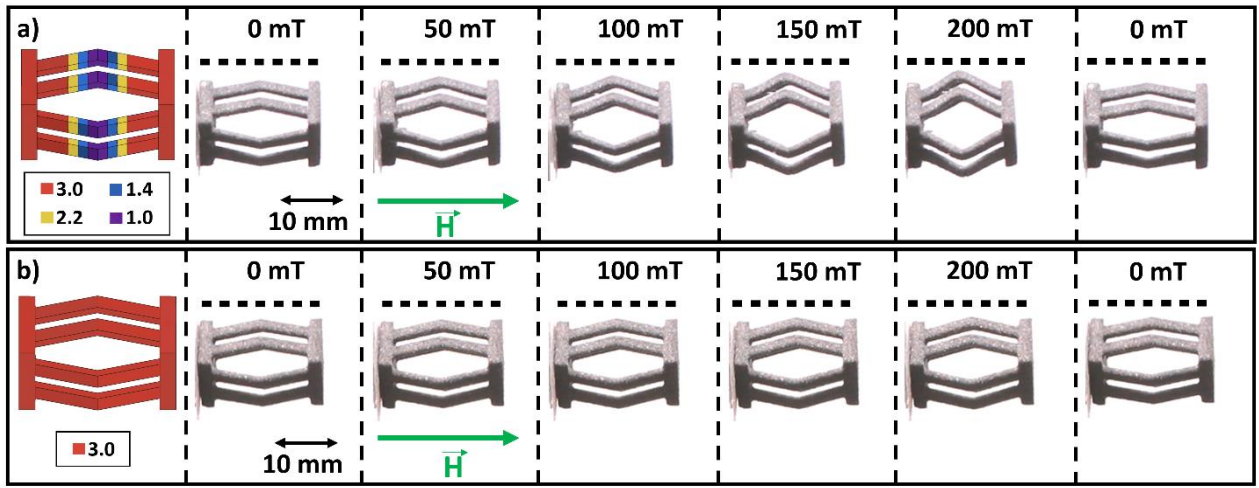

*Figure 9: Comparison of artificial muscle shape beam as a function of the magnetic actuation field. a) presents the shape at different external magnetic actuation fields which was produced with lower energy scale parameter in the middle as indicated by the colors, while b) is produced homogeneously with the same process parameters.*

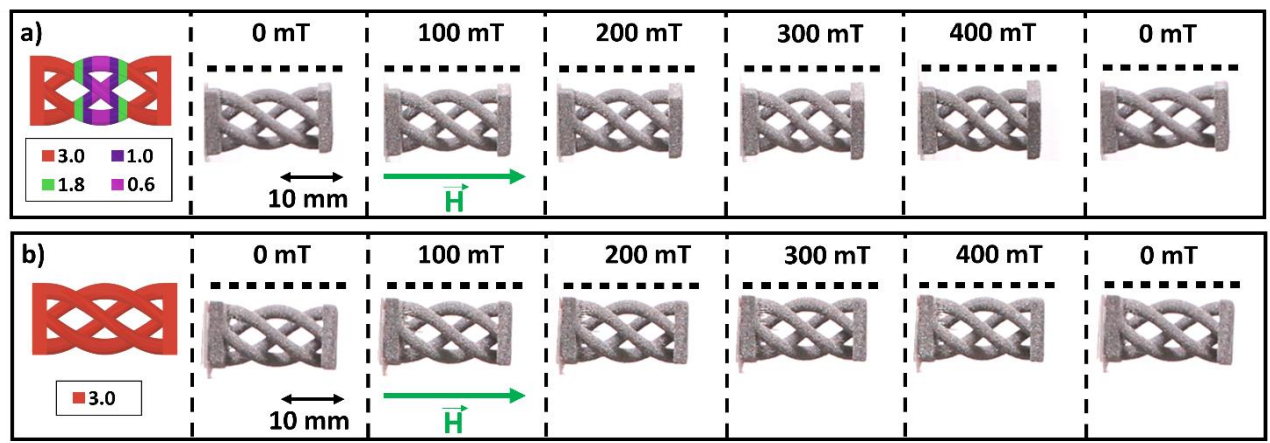

*Figure 10: Comparison of artificial muscles with a helix shape as a function of the magnetic actuation field H. a) presents the shape at different external magnetic actuation fields which was produced with lower energy scale parameter in the middle as indicated by the colors, while b) is produced homogeneously with the same process parameters.*

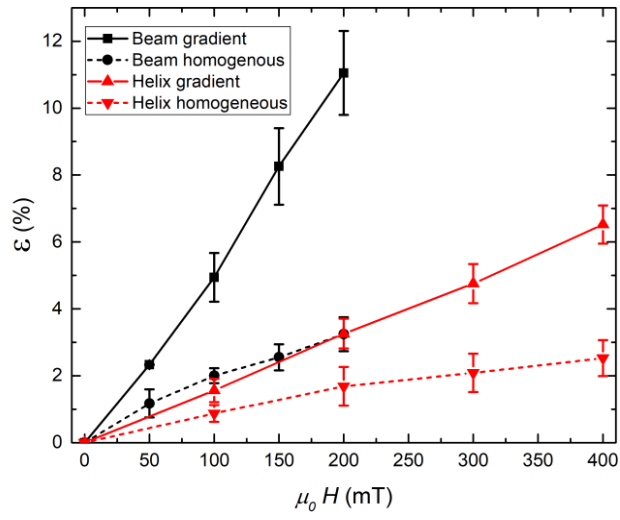

*Figure 11: Deformation ε as function of the external magnetic actuation field H for the two different artificial muscle shapes beam and helix with and without locally adjusted properties.*

### 3.4.2 Foldable cube

Another example of an application for the locally adjusted magneto-active composites is a foldable cube. The hinges can be made soft while the faces of the cube remain stiffer. The idea is to allow higher deformation where it is needed and remain stiffer where higher structural integrity is required. This can be utilized in assembly applications including origami techniques [64].

Figure 12 compares the shape of two cubes with four sides at different external magnetic actuation fields H, processed once with locally adjusted energy scale process parameters Figure 12 a) and once with no process parameter variation [Figure 12 b)]. The different energy scale parameters are indicated in the figures by the different colors. As can be seen, the locally adjusted cube reaches a larger degree of deformation at the same actuation field. While the transformation into the cube shape is almost completed at 100 mT for the sample with modulated stiffness, the transformation of the homogenous shape is not completed at 300 mT. The factor three lower actuation field of the stiffness modulated cube helps to reduce the energy consumption in an application.

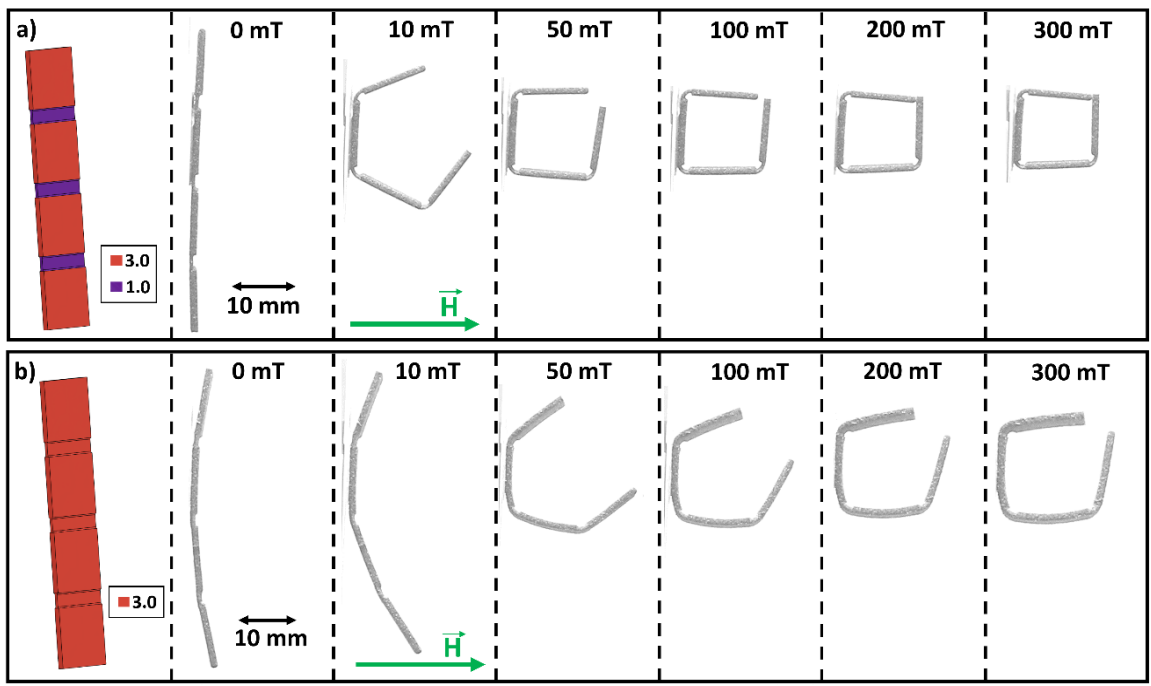

*Figure 12: Comparison of magnetically actuated folding cubes at different magnetic actuation fields H. a) Cube with soft hinges between the sides, which allows higher degree of folding at the same magnetic actuation fields in comparison to b) where the sides and the hinges are produced with the same process parameters.*

The cubes with four segments are magnetized in a folded state [Figure 13 (a)]. If the cube is unfolded again, a specific magnetization profile is imprinted [Figure 13 (b)]. Upon applying an external magnetic field to the unfolded cube, there is a driving force back into the folded state. To verify that the different actuation response results from the different mechanical properties and not from a different magnetization profile, the imprinted magnetization profile was indirectly measured with our custom build stray field scanner. The magnetic stray field close to the surface (1 mm) of the composite indicates the magnetization direction of the composite. The stray field as a function of the position above the cube shapes is presented in Figure 13 for the locally adjusted shape in c) and for the homogenous shape in d). The values are normalized to one, only the direction of the stray field is presented. The stray field follows the imprinted magnetization profile and no significant difference can be observed between the two different samples.

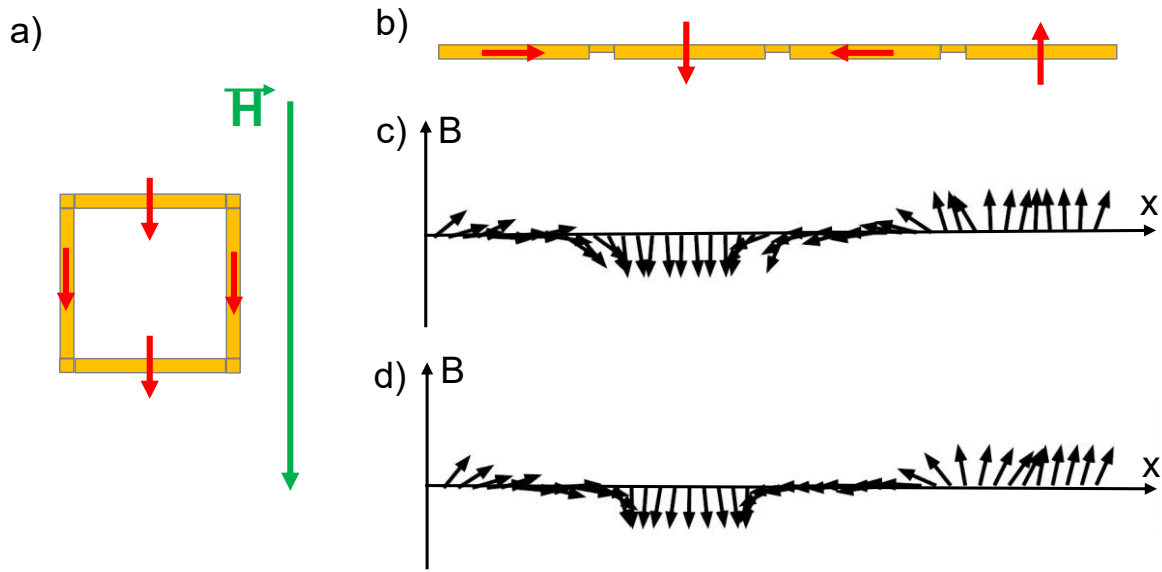

*Figure 13: a) Schematic representation from the side of magnetisation profile of a cube in folded state during magnetization and in b) unfolded state as a line. The measure magnetic flux density B, measured with the stray field scanner 1 mm above the surface of the cubes, which indicates the magnetization direction, is presented in c) for the graded sample and d) shows the flux density profile for the homogenous sample.*

### 3.4.3 Airway stent

Magneto-active composites are already explored for biomedical applications [12–14]. The possibility to tailor geometry and stiffness simultaneously is highly important for biomedical applications, where actuators work in confined spaces, which results in limited design freedom. One example is an airway stent, which is used to treat airway obstructions. The placement of the stent serves the purpose of preserving the openness of the airway, effectively counteracting the constriction caused by tumors, lymph nodes, scar strictures, or sealing cavities [65]. Preferred are individualized stents, specifically designed for the patient´s anatomy and disease [66]. At the area of the external compression by e.g. a tumor should be accompanied by a greater expansion force at this location, generated by an increased stiffness at this area of the stent. On the other hand, for sealing a cavity the stent should be very soft and flexible around this area [66]. Here we propose a a magnetically controllable airway stent inside a representation of a trachea with a schematic tumor, Figure 14 a). The trachea, also referred to as the windpipe, is a tubular structure in the human body responsible for transporting air from the throat and larynx to the bronchi. The typical dimensions for the trachea in males are 10 - 13 cm in length and 1.8 cm in diameter [67]. The stents presented here were magnetized before in a contracted state, as shown in Figure A 6. Therefore, the stent contracts upon the application of a magnetic field (500 mT) as visible in Figure 14 b) and can be delivered to the desired location with the field (Figure 14 c)). Due to the contracted state in which the outer diameter of the stent is reduced by 30 %, the stent can be delivered minimally invasive in surgery. At the target location next to the tumor (Figure 14 d)), the magnetic field is switched off, which results in the expansion of the shape to counteract the contraction forces of the airway obstruction as visible in Figure 14 e). The corresponding video of the actuation demonstration is in supplementary V1. Since the stent is presented in a schematic trachea in an open cylinder, the stent can not generate large forces on the tumor. Therefore the same experiment was performed with a cylinder where only one-quarter of the circumference is missing. A comparison of a stent that was produced with homogenous process parameters and a locally adjusted stent is presented in Figure 14 f) and g), respectively. Only the end position after the removal of the magnetic field is presented. The corresponding videos are in supplementary V2 and V3. While the homogenously produced stent has not enough stiffness to push out the tumor, the locally adjusted stent, which provides a higher stiffness at the position of the tumor can push out the tumor almost completely. After treatment accomplishment the stent could be easily magnetically repositioned or removed by the application of a magntic field of 500 mT. For an application in biological systems, the stent must be covered with a biocompatible coating for example silicones or hydrogels.

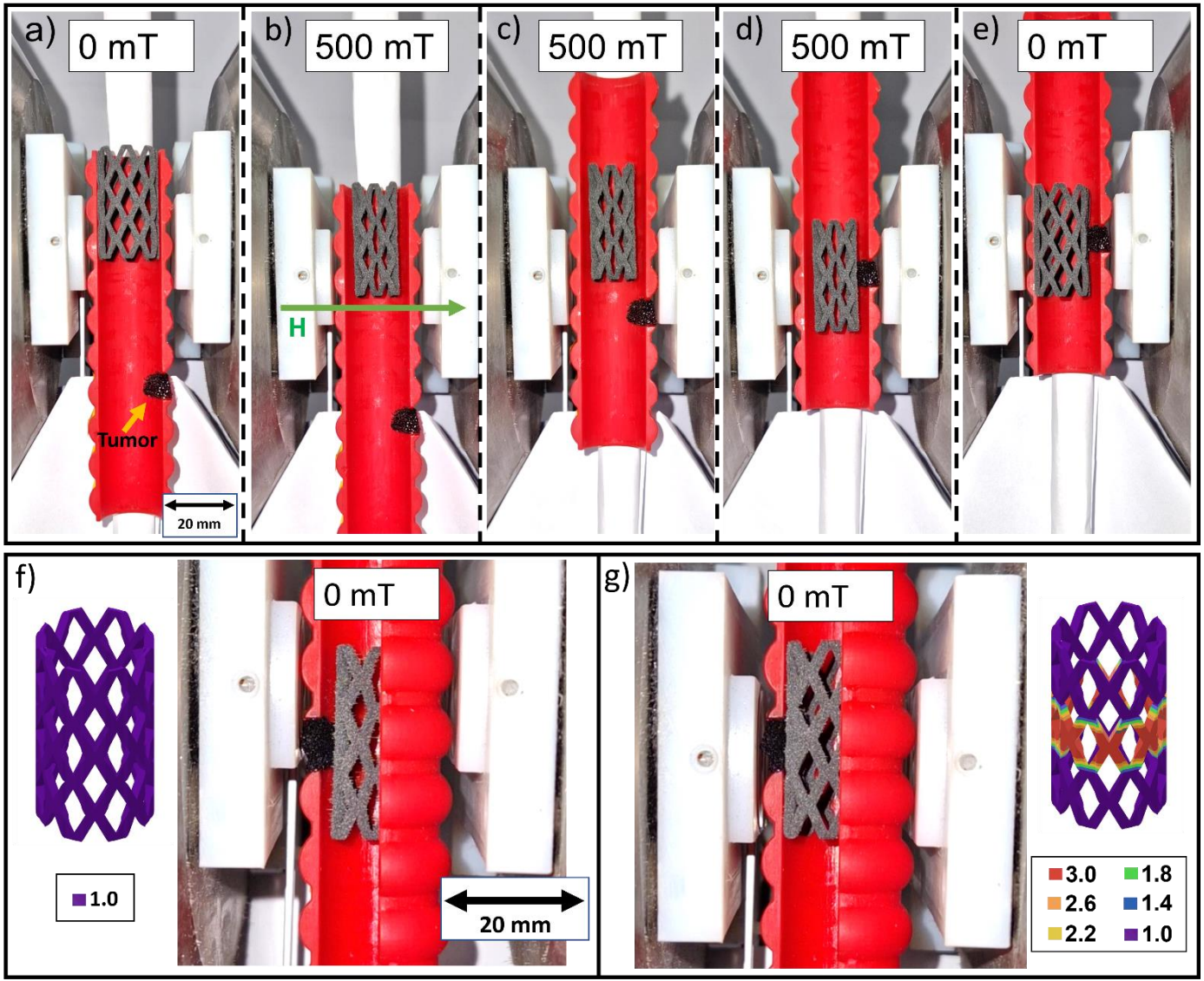

*Figure 14: a) Stent shape in the trachea with tumor. b) By applying a magnetic field of 500 mT, the stent diameter undergoes contraction, and the positioning can be controlled based on the magnetic field's location as indicated in c). d) The stent is moved to the target position at the location of the tumor. e) With the removal of the magnetic field, the stent reversible transforms into the original shape with a larger diameter, creating an expansion force on the tumor. Comparison of the diameter of the stent after field removal at the tumor position. The homogeneously produced stent creates a smaller expansion force at the tumor f) in comparison to the locally adjusted stent, which is stiffer in the center as shown in g).*

## 4. Conclusions and Outlook

The aim of this study was to develop magneto-active materials with locally varying stiffness by LPBF using commercially available machines, materials and locally adjusting the energy scale process parameter. For local adjustment, the digital files of the samples must be divided as follows: in the x-y plane, the sections of the digital files must possess the same dimensions as the hatch distance. For local adjustment in z-direction, parallel to the build-up direction, the digital files must have dimensions in z-direction which are integer multipliers of the layer height. The minimum step size where the properties can be tailored are therefore the layer height in z-direction and the hatch distance in the x-y plane. This method was validated with a composite of TPU polymer matrix and hard magnetic atomized Nd-Fe-B particles. The range of processing parameters was determined by density, magnetic and mechanical property characterization. The change of the laser parameters results in a different level of porosity, ranging from 2% to 24% for decreasing laser power, affecting the magnetic properties. For low laser powers, the presence of high porosity leads to magnetic particles that are weakly embedded, causing them to rotate during the demagnetization process. Since this effect is limiting the actuation performance, the corresponding process parameter (0.6) is here the lower boundary condition. Tensile tests showed a range of Young's moduli between 2 and 22 MPa. The gradient in mechanical properties was verified by a sequence of local hardness measurements on samples printed with different orientations towards the build-up direction.

The actuation behavior of magneto-responsive actuators with locally tailored stiffness was then studied under a magnetic field. Importantly, higher deformation at the same field or enhanced deformation at lower magnetic fields in comparison to their homogenous counterparts is observed. That the actuation can be completed with lower actuation field results in an increase the energy efficiency of a system. Also, the presented actuators were tailored specifically for an application where the material properties can be adjusted to the requirements. One biomedical example is a magnetically actuated airway stent, where the location of the contracted stent within the airway can be controlled by the field and the removal of the field results in the expansion of the diameter at the desired location. Here the stiffness was adjusted to increase the expansion forces at the site of an external compression.

Next steps are the production of highly porous structures with a strong filler matrix adhesion to prevent particle rotation to expand the range of tailorable stiffness to lower values. The usage of non-spherical particles can help to tackle this challenge. Simultanous optimization of the geometry and the materials property by simulation will help to design actuators and parts specifically for applications while mitigating contrains in their weight, maximal dimensions and required actuation fields.

# Appendix

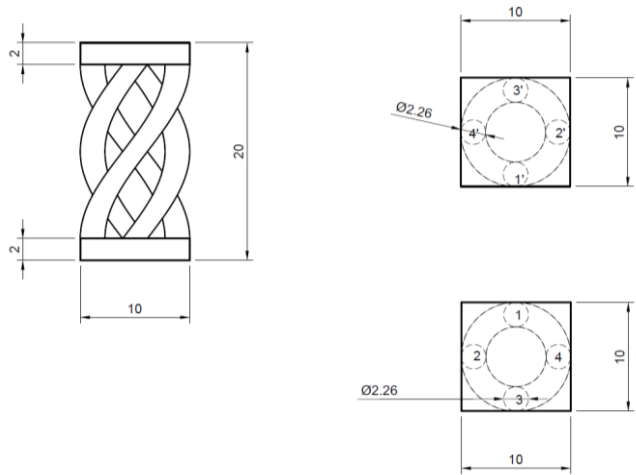

*Figure A 1: Dimensions of the helix actuator in mm.*

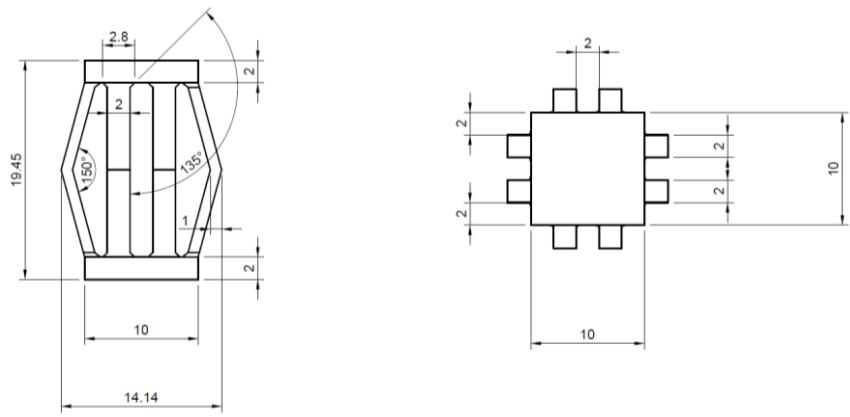

*Figure A 2: Dimensions of the beam actuator in mm.*

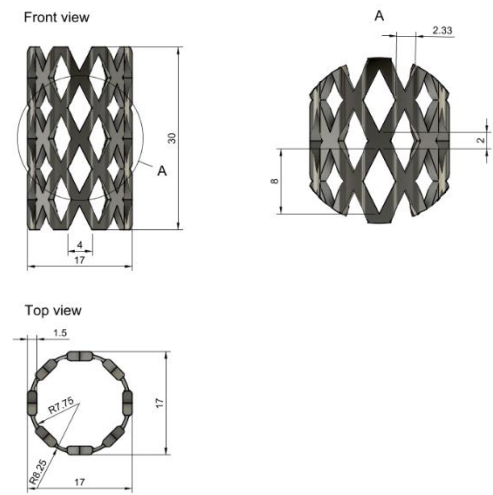

*Figure A 3: Dimensions of the stent actuator in mm.*

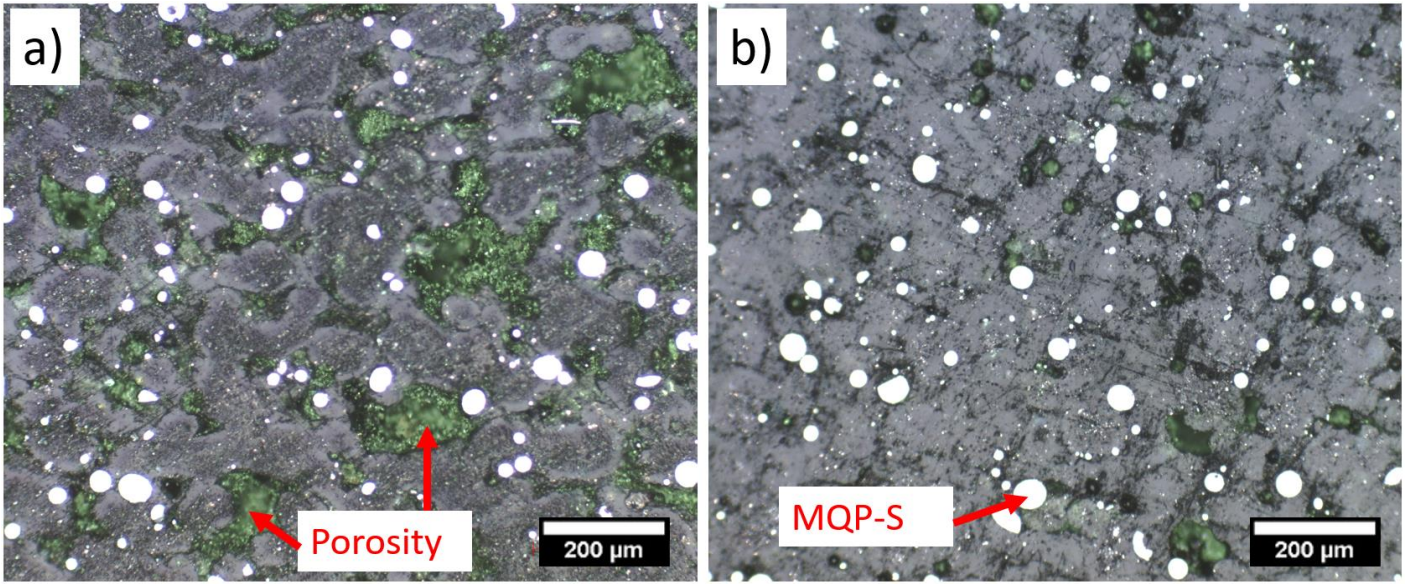

*Figure A 4: Light microscopy image of the composites fabricated with an energy scale value of 0.6 is shown in a) and in b) 3.0. The distribution of the MQP-S particles inside the TPU polymer is visible, as well as a high porosity in the 0.6, in comparison to the 3.0 sample, which has a significant lower porosity. The very small particles are Cu particles, which are occur during the grinding process, since the sample is embedded in a Cu containing polymer matrix.*

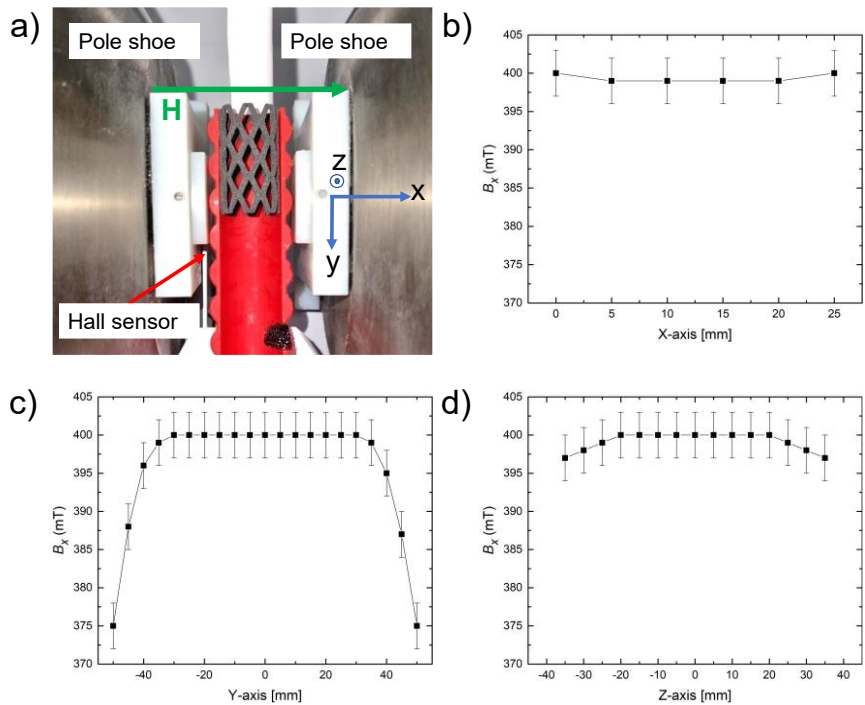

*Figure A 5: Magnetic actuation setup in VSM at nominal 400 mT. Bx Magnetic field component in x-direction a), z-direction b) and y-direction d). Measurement error is 3 mT.*

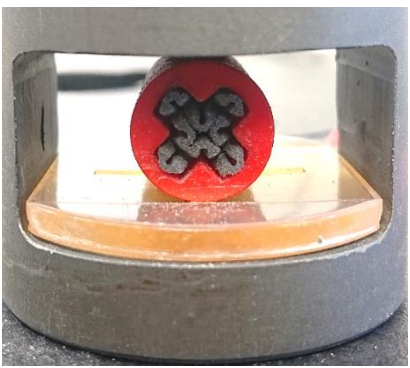

*Figure A 6: Stent in comressed shape during magnetization*

Table A.1: Geometric density and the standard deviation Δ of the geometric density, remanence $B_r$ and coercivity $H_c$ of the composites produced with different energy scale parameters.

| Energy Scale | Density (g/cm³) | ΔDensity (g/cm³) | Porosity (%) | $B_r$ (mT) | $H_c$ (mT) |
|---|---|---|---|---|---|
| 0.6 | 1.09 | 0.04 | 24.25 | 47 | 836 |
| 1.0 | 1.19 | 0.04 | 17.59 | 61 | 873 |
| 1.4 | 1.21 | 0.04 | 16.08 | 66 | 882 |
| 1.8 | 1.28 | 0.03 | 11.04 | 69 | 884 |
| 2.2 | 1.33 | 0.03 | 7.42 | 72 | 885 |
| 2.6 | 1.38 | 0.02 | 4.33 | 74 | 888 |
| 3.0 | 1.41 | 0.03 | 2.39 | 76 | 885 |

Table A.2: Calculated average values and their corresponding standard deviations **Δ** of the characteristic values obtained from the engineering stress-strain curves of the uniaxial tensile tests. $\sigma_b$ is the tensile strength at break. $\varepsilon_m$ and $\varepsilon_b$ the elongation at strength and elongation at break, respectively.

| Energy Scale | E (MPa) | ΔE (MPa) | $\sigma_m$ (MPa) | $\Delta\sigma_m$ (MPa) | $\sigma_b$ (MPa) | $\Delta\sigma_b$ (MPa) | $\varepsilon_m$ (%) | $\Delta\varepsilon_m$ (%) | $\varepsilon_b$ (%) | $\Delta\varepsilon_b$ (%) |
|---|---|---|---|---|---|---|---|---|---|---|
| 0.6 | 2.22 | 0.40 | 0.07 | 0.01 | 0.06 | 0.01 | 18.25 | 2.27 | 20.91 | 2.58 |
| 1.0 | 6.72 | 0.79 | 0.28 | 0.04 | 0.24 | 0.04 | 27.79 | 3.34 | 29.61 | 3.79 |
| 1.4 | 8.83 | 0.55 | 0.44 | 0.05 | 0.41 | 0.05 | 35.17 | 3.27 | 36.56 | 3.16 |
| 1.8 | 13.11 | 0.98 | 0.62 | 0.05 | 0.57 | 0.06 | 41.99 | 2.61 | 44.22 | 2.87 |
| 2.2 | 16.85 | 2.09 | 0.73 | 0.06 | 0.67 | 0.06 | 36.67 | 2.72 | 37.73 | 2.96 |
| 2.6 | 19.92 | 2.54 | 0.86 | 0.10 | 0.79 | 0.07 | 36.90 | 3.74 | 39.63 | 4.35 |
| 3.0 | 21.99 | 1.59 | 0.99 | 0.07 | 0.92 | 0.07 | 36.64 | 4.16 | 38.88 | 4.52 |

**Acknowledgement**

This work was financially supported by the Deutsche Forschungsgemeinschaft (DFG, German Research Foundation), Project ID No. 405553726, TRR 270 and the RTG 2761 LokoAssist (Grant no. 450821862). The authors thank Semih Ener for the fruitful discussions and Bernd Stoll for his help with the development of the strayfield scanner. A part of the graphical abstract was created with BioRender.com.

**Data Availability Statement**

Data available on request from the authors.

**Declaration of Competing Interest**

None.